\documentclass{article}

\usepackage{graphicx}
\usepackage{amsmath}
\usepackage{amssymb}

\usepackage{subcaption}
\begin{document}
\noindent {\bf Formal derivation of an inversion formula for the approximation of interface defects by means of active thermography}
\\
\\
{Gabriele Inglese$^1$ (orcid: 0000-0002-5054-0713) and Raffaele Inglese$^2$ }
\\




$^{1}$ CNR-IAC Sesto Fiorentino, Italy; gabriele@fi.iac.cnr.it\\
$^{2}$ Scuola di Ingegneria Universita' della Basilicata, Potenza, Italy; raffaele.inglese@unibas.it
\\
\\





\abstract{Thermal properties of a two-layered composite conductor are  modified in case the interface  is damaged. The present paper deals with nondestructive evaluation of perturbations of interface thermal conductance due to the presence of defects. The specimen is heated by means of a lamp system or a laser while  its surface temperature is measured with an infrared camera in  the typical framework of Active Thermography. Defects affecting the interface are evaluated using an inversion formula obtained  by means of Laplace transformation and  suitable symmetries of parabolic differential operators (reciprocity). Results of numerical inversion from simulated data are encouraging.}
\\ \\
\section{Introduction}

Thermal properties of a two-layers composite conductor are  modified in case the interface  has been damaged. The present paper deals with nondestructive evaluation of perturbations of the thermal contact conductance (TCC) of the interface due to the presence of defects. The specimen is heated by means of a lamp system or a laser while  its surface temperature is measured with an infrared camera in  the typical framework of Active Thermography   \cite{Ma01} represented schematically in Figure .  Mathematical model consists of a system of two Initial Boundary Value Problems (IBVPs) for heat equation.  The strategy for evaluating the interface is borrowed from applications of reciprocity techniques to the solution of inverse problems (see for example \cite{ABC20,AB93,FCA23}) but our main reference is \cite{BAB03}. \\ 
An approximated formula for evaluating the perturbation of TCC is derived in section \ref{sec:pert} and tested numerically in section \ref{sec:numer}.

\subsection{Layered domains: Thermal Contact Conductance of interfaces}
Consider a composite body made up of two thermally conducting layers divided by a very thin interspace  
filled up with air or other poorly conductive materials. As long as the specimen is heated by an external source, heat flows through the interspace mainly in correspondence to possible contact spots between the conducting layers.  We assume that the effect of the interspace on  heat transfer between the two layers  is  correctly modeled in terms of transmission conditions  on an interface  $\Sigma$ (of codimension $1$) that separates the conducting layers. 
Interfaces can be classified as perfect or imperfect according to their thermal properties. Here, we deal with  a {\it Low Conductivity Imperfect } (LCI) interface   which allows for a temperature jump with continuous heat flux \cite{JKS14}.

The  {\it Thermal Contact Conductance} (TCC) $h$ of an interface $\Sigma$ is a non-negative parameter defined as the  absolute value of the ratio between the heat flux through $\Sigma$ to the temperature jump across it.   TCC is useful in mathematical modeling of heat transfer through interfaces because it becomes the exchange coefficient in the normal form of 3rd kind (Robin) transmission conditions. Its inverse $r=\frac{1}{h}$ (as long as $h>0$) is referred to as {\it Thermal Contact  Resistance} (TCR) (see for example \cite{IDBL03} Ch 3). In the limit  case $h \to 0$ the interface is perfectly insulating.

In this paper, we focus on a  framework  in which the undamaged interface has a known constant TCC $h_0>0$ and the defect produces a local perturbation of TCC described by a function $\delta h > 0$ (insulation degradation) or  $\delta h < 0$ (delamination i.e. more insulating interface).   The mathematical model where a non constant TCC  is the exchange coefficient in Robin transmission conditions (see (\ref{TC21}) and (\ref{TC22}) in section \ref{sec:geom}), though not rigorously founded, is in agreement with experiments  and  it is successfully used among practitioners (see for example \cite{ACOA16,SBecc22}). 

\subsection{The direct model and the inverse problem}\label{sec:dirinv}

In this section, we describe briefly the mathematical model and the approach used here to solve the inverse problem. The lower layer $\Omega^-$ is heated by means of thermal flux coming from below e.g. by a lamp kept on for a time interval of $t_{ON}$ seconds. Heat passes through the interface $\Sigma$ of TCC equal to $h$  so that the temperature of $\Omega^+$ changes in time. Heat transfer through $\Sigma$   is modeled by means of Robin transmission boundary conditions
(see for example \cite{PP09, CGS18}). 

If $h$ is given, the temperature of $\Omega$ is the unique solution of an Initial Boundary Value Problem for heat equation (direct model).

In order to identify an unknown $\delta h=h-h_0$ (inverse problem), a   sequence $\psi$ of temperature maps  is taken on the top surface of  $\Omega^+$ in the time interval $(0,t_{max})$.  This setting is usually referred to as {\it transmission mode} in Long Pulse Thermography  \cite{MCG04}.  

We assume that the (background)  temperature of the undamaged specimen ($h=h_0$) heated up by $\phi$  is fully known.    \\ 

It is remarkable that $h$ is independent of time (at least in the time scale of $t_{max}$  with $t_{max}$ quite bigger than $t_{ON}$) so that  Laplace transform can be applied to equations and boundary conditions (see section \ref{sec:trans}) and no inverse transform is required. In this way we obtain a system of two BVPs   for elliptic equations in  $\Omega^+$ and ${\Omega}^-$ (connected by Robin transmission conditions) whose solutions $U^+$ and $U^-$ are the Laplace transform of the temperatures of the two layers. In \cite{SBecc22, IO23} we approach  the inverse problem by means of perturbation theory and Thin Plate Approximation.  An inversion method based on the reciprocity principle is described and tested in \cite{FCA23}. In  \cite{FCA23},  the test functions that characterize the method come from the numerical solution of suitable boundary value problems. Here we use reciprocity with  a choice of explicit test functions  (\ref{ftest}) borrowed from \cite{BAB03}. A linear relation (see (\ref{claim}))  between the (Laplace transformed) temperature jump at $\Sigma$ and the TCC $h$  is extrapolated from one dimensional theory.   Hence, we derive a formal relation which (after a suitable choice of Laplace frequency parameter) allows us to compute cosine Fourier coefficients of $\delta h(x)$ (see (\ref{rec01del})). 

\section{Geometry in 2D, notation, direct  model and inverse problem}\label{sec:geom}

\begin{figure}[h] 
    \centering 
    \includegraphics[width=0.5\textwidth]{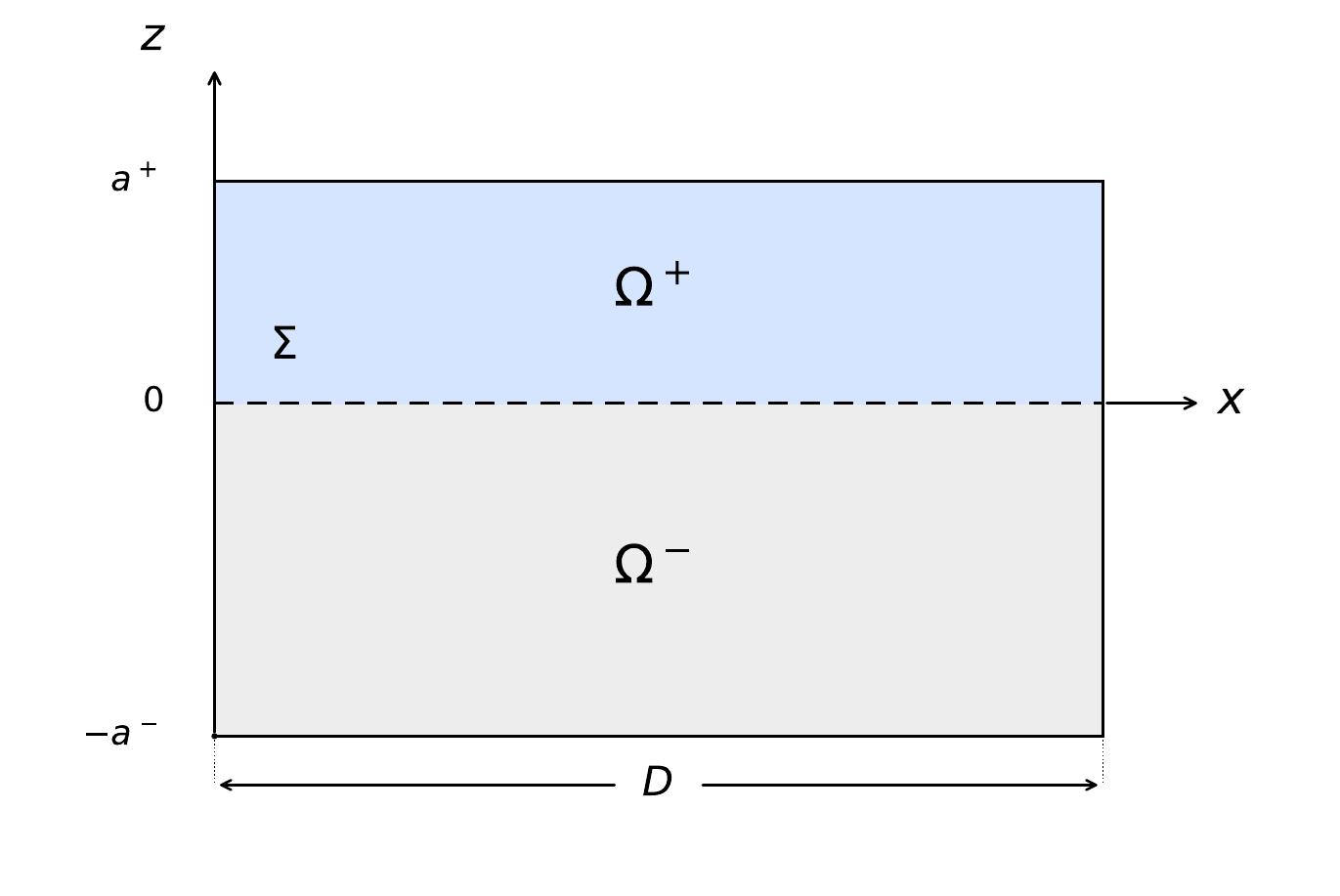} 
    \caption{Geometry of the problem in 2D:  $\Omega=\Omega^- \cup \Sigma \cup \Omega^+$}
    \label{fig:domain} 
\end{figure}

Here we deal with a problem in two space dimensions (see figure \ref{fig:domain}) . This assumption lightens the notation without consequences on generality.\\

\noindent Let $\Omega$ be the rectangle $(0,D) \times  (-a^-,a^+)$  in the 2D space $(x,z)$.  

\noindent Let $\Omega^+$ be $(0,D) \times  (0,a^+)$ and  $\Omega^-$ be $(0,D)\times  (-a^-,0)$. 

\noindent Let $\Sigma=\{ x \in (0,D) \ , \ z=0\}$. Clearly $\Omega= \Omega^+ \cup  \Sigma \cup \Omega^-$.\\

The inverse  problem at hand is closely related to the class of Inverse Heat Conduction Problems that are well known to be severely ill-posed (see \cite{Back23}). Hence the geometrical assumption that $\frac{a^++a^-}{D}<1$. 

%

Thermal behavior of each layer $ \Omega^\pm$ is determined by its conductivity $ \kappa^\pm$, density $\rho^\pm$ and  specific heat $c^\pm$. 

Let $u^\pm(x,z,t)$ with $(x,z)\in  \Omega^\pm$ and $t>0$ the temperature increase (with respect to an initial and surrounding temperature $U^0$) in $ \Omega^\pm$ obtained by applying, for a time interval $(0,t_{ON})$, a heat flux $ \phi(x,t)$ to $ \Omega^- $ (more precisely, $ \phi(x,t)=0$ for $t >t_{max} $) .   Clearly, $u^\pm(x,z,0)=0$. Assume that the vertical sides of the composite domain are insulated while the horizontal sides exchange heat with the environment. The thermal contact conductances of top ($z=a^+$) and bottom side ($z=-a^-$) are the positive constants $ h^+$ and $ h^-$ respectively.

\noindent {\bf Notation} $f_t, f_x, f_z$ are partial derivatives; $f_\nu$ means   ``outward normal derivative" of $f$ with respect to a boundary which is unambiguous in the context;
$\int_{\partial E} f d\gamma$ is the line integral of $f$ on the curve $\partial E$. 

\subsection{The Direct Model}\label{dirmod} 
Given the constant parameters $a^\pm$, $ \kappa^\pm$, $\rho^\pm$ , $c^\pm$ and $h^\pm$  and  given the interface thermal conductance $ h(x)$,  the functions  $u^\pm$ fulfill the following Initial Boundary Value Problem (IBVP) for the heat equation in the composite domain $ \Omega$ :\\

\noindent $IBVP^-$

\begin{equation}\label{equatm}
 \rho^-  c^- u^-_t =  \kappa^-  (u^-_{xx}+ u^-_{zz})\ , \  (x,z) \in \Omega^- , \  t>0
\end{equation}
\begin{equation}
- \kappa^- u^-_z(x,-a^-,t)+ h^- u^-(x,-a^-,t)=\phi(x,t)
\end{equation}
($u^-_\nu=0$ on the {\it vertical} sides of $\Omega^-$)\\

\noindent $IBVP^+$
\begin{equation}\label{equatp} 
 \rho^+  c^+ u^+_t =  \kappa^+  (u^+_{xx}+u^+_{zz}) \ , \  (x,z) \in \Omega^+ , \  t>0
\end{equation}
\begin{equation}
\kappa^+ u^+_z(x,a^+,t)+h^+u^+(x,a^+,t)=0
\end{equation}
($u^+_\nu=0$ on the {\it vertical} sides of $\Omega^+$)\\

\noindent with {\it transmission conditions}\\

\begin{equation}\label{TC21}
 \kappa^- u^-_z(x,0,t)+h(x)[u](x,t)=0
 \end{equation}
where $[u](x,t)=u^-(x,0,t)-u^+(x,0,t)$,  and
\begin{equation}\label{TC22}
\kappa^- u^-_z(x,0,t)=\kappa^+ u^+_z(x,0,t).
\end{equation}\\

Initial data are
\begin{equation}\label{init}
u^-(x,z,0)=0 \ , \  (x,z) \in \Omega^- 
\end{equation}
\begin{equation}
u^+(x,z,0)=0  \ , \  (x,z) \in \Omega^+.
\end{equation}\\

\noindent {\it Remark} If $\phi$ and $h$ are continuous functions and $H^1(\Omega)$ is a product Hilbert space equipped with a suitable norm, the system (\ref{equatm})-(\ref{init}) admits for all $T>0$ a unique solution $(u^-,u^+) \in L^2(0,T;H^1(\Omega))$, stable with respect to  error on $h$ (see \cite{IO22}).

\subsection{The Interface Inverse Problem}\label{sec:inverse}  Let $h(x)=h_0+\delta h(x)$ be the unknown perturbed TCC with given $h_0$. 
Our goal is to approximate $\delta h(x)$ from the knowledge of:

1) the controlled flux $\phi$ (a source heats up the bottom surface of $\Omega$ for $t_{ON}$ seconds);

2) an additional (boundary) dataset  $ \psi(x,t)=u^+(x,a^+,t)$ for $t\in (0,t_{max})$ (the temperature of the top surface of $\Omega$ taken by  a thermal imaging camera).

section{Laplace transform of the direct problem}\label{sec:trans}

We know that the bounded functions $u^\pm(x,0,t)$ that describe the temperature increase of $\Omega^\pm$ at time $t$ are  decreasing for $t>t_{max}+\delta t$   ($\delta t \ge 0$  depends on thickness and diffusivity of the specimen).  Hence, the temperature data $\psi(x,t)$ can be extended  for $t>t_{max}$ to  bounded functions $\psi_\infty$ that decrease to zero for $t \to \infty$.\\

Define (for all real positive numbers $A$) the Laplace transform of $u^\pm(x,z,t)$ and $\psi$  as
\begin{equation}
{U_A}^\pm (x,z) =  \int_0^\infty u^\pm(x,z,t)e^{-A t}dt
\end{equation} 
and
$$\Psi_A(x) = \int_0^\infty \psi_\infty(x,t)e^{-A t}dt.$$
In the same way (assigning for completeness $\phi(x,t)=0$ for $t >t_{max}$):
$$\Phi_A(x) = \int_0^\infty \phi(x,t)e^{-A t}dt.$$\\

In what follows ${U_A}^\pm$, $\Psi_A$ and $\Phi_A$ are written  simply ${U}^\pm$, $\Psi$ and $\Phi$.  This does not mean that the choice of $A$ is a minor aspect of gthr problem. It will be discussed in section \ref{sec:stab} about stability of reconstruction formula.
Hence, standard calculations change  (\ref{equatm})-(\ref{init}) into the following system of elliptic BVPs\\

\noindent $BVP^-$

\begin{equation}\label{equatmL}
 \rho^-  c^- A {U}^-= \kappa^-({U}^-_{xx}+{U}^-_{zz})\ , \  (x,z) \in \Omega^- , 
\end{equation}
\begin{equation}
-\kappa^- {U}^-_z(x,-a^-)+  h^- {U}^-(x,-a^-)=\Phi(x)
\end{equation}
(${U}^-_\nu=0$ on the {\it vertical} sides of $\Omega^-$)\\

\noindent $BVP^+$
\begin{equation}\label{equatpL} 
 \rho^+  c^+A {U}^+ =  \kappa^+ ({U}^+_{xx}+ {U}^+_{zz}) \ , \  (x,y,z) \in \Omega^+ ,
\end{equation}
\begin{equation}
\kappa^+ {U}^+_z(x,a^+)+h^+{U}^+(x,a^+)=0
\end{equation}
(${U}^+_\nu=0$ on the {\it vertical} sides of $\Omega^+$) 
with {\it transmission conditions}
\begin{equation}\label{TC1L}
 \kappa^- {U}^-_z(x,0)+ h(x)[U](x,0)=0
\end{equation}
and
\begin{equation}\label{TC2L}
\kappa^- {U}^-_z(x,0)=\kappa^+ {U}^+_z(x,0)
\end{equation}
where 
\begin{equation}\label{diffe}
[U](x)={U}^-(x,0)-{U}^+(x,0).
\end{equation}

\section{Reciprocity conditions}\label{sec:recgap}

\noindent Let $v=(v^-,v^+)$ be a solution of the backward heat conduction problems
\begin{equation}\label{back}
 \rho^\pm  c^\pm v^\pm_t(x, z, t) + \kappa^\pm \Delta v^\pm(x, z, t)=0
\end{equation}
in $\Omega^\pm \times (0,T)$ with 
\begin{equation}\label{lim0}
\lim_{T \to \infty} v\pm(x, z, T) = 0
\end{equation}
uniformly in $\Omega$.  Additional requirements are
\begin{equation}
v_x^\pm(0,z,t)=v_x^\pm(D,z,t)=0
\end{equation}
and
\begin{equation}
\kappa^-v_z^-(x,0^-,t)=v_z^+(x,0^+,t)
\end{equation}
where $v_z^-$ is a left derivative and $v_z^+$ a right one.
In what follows we refer to $v$ as to a {\it test function}. Assume also that $\sup_{\partial \Omega} |\nabla v| $ is uniformly bounded for all $T >0$. \\
 
Let  $(u_0^-,u_0^+)$ be  the solution of   (\ref{equatm})-(\ref{init})  when $h \equiv h_0$
i.e. the background temperature increase  in $\Omega^+ \cup \Omega^- \times (0,T)$ while $(u^-,u^+)$ is the solution corresponding to $h=h_0+\delta h$  (the perturbed value of TCC due to a defect in the interface).  As already noted in section \ref{sec:dirinv}, we are assuming that  $(u_0^-,u_0^+)$ is known.

For all $T > 0$, we have the pair of equations
\begin{equation}\label{green}
\int_0^T \int_{\Omega^\pm} (v^\pm\Delta u^\pm - u^\pm\Delta v^\pm)  dxdzdt = \int_0^T  \int_{\partial \Omega^\pm } (v^\pm u^\pm_\nu - u^\pm v^\pm_\nu ) d\gamma dt .
\end{equation}\\

On the other hand, since $u^\pm(x,z,0)=0$, we have  also
\begin{equation}\label{lim2}
\begin{split}
\int_0^T  \int_{\Omega^\pm}(v^\pm \Delta u^\pm - u^\pm \Delta v^\pm) dx dzdt =
& \int_{\Omega^\pm}  \int_0^T \frac {v^\pm u^\pm_t+u^\pm v^\pm_t}{\alpha^\pm} dt dx dz \\
= & \frac{1}{\alpha^\pm} \int_{\Omega^\pm}  u^\pm(x,z,T)v^\pm(x,z,T)dx dz=r^\pm(T).
\end{split}
\end{equation}\\

It follows from (\ref{lim0}) that $\forall \epsilon>0$  $\exists {\bar T}$ such that  $|r^\pm(T) |< \epsilon$ if $T >{\bar T}$.  Roughly speaking, the l.h.s. of (\ref{lim2}) is $\approx 0$ if $T$ is large enough.\\

It comes  from (\ref{green}), (\ref{lim2})  that for $T$ large enough, the following {\it reciprocity conditions}  holds separately in $\Omega^+$ and $\Omega^-$
\begin{equation}\label{reconmeno}
 \int_0^T \int_{\partial \Omega^- \setminus \Sigma} (v^-u^-_\nu - u^- v^-_\nu)d\gamma dt+\int_0^T\int_{\Sigma} (v^- u^-_{\nu}- u^-v^-_{\nu} )d\gamma dt \approx  0
\end{equation}
\begin{equation}\label{reconpiu}
 \int_0^T \int_{\partial \Omega^+\setminus \Sigma} (v^+u^+_\nu - u^+ v^+_\nu)d\gamma dt+\int_0^T\int_{\Sigma} (v^+u^+_{\nu}-u^+v^+_{\nu} )d\gamma dt   \approx  0
\end{equation}

\noindent {\bf Notation:} In order to lighten formulas up to (\ref{rec01}) it is  $f(z) \equiv f(x,z,t)$ ( the variable $z$ only is written explicitly).\\

Since normal derivatives of $u^\pm$ and $v^\pm$ vanish at the vertical sides of $\Omega$, equations (\ref{reconmeno}) and (\ref{reconpiu})  become

\begin{equation}\label{recon2meno}
\kappa^- \int_0^T\int_{\Sigma} (v^- u^-_{z}(0) -u^- v^-_{z}(0) )d\gamma dt \approx \kappa^- \int_0^T \int_{0}^D (v^-u^-_z(-a^-) -u^- v^-_z(-a^-))d\gamma dt
\end{equation}
\begin{equation}\label{recon2piu}
\kappa^+\int_0^T\int_{\Sigma} ( v^+u^+_{z}(0)-u^+v^+_{z}(0))d\gamma dt  \approx  \kappa^+\int_0^T \int_{0}^D (v^+u^+_z(a^+) - u^+ v^+_z(a^+))d\gamma dt
\end{equation}

Observe that the terms on the right hand side of (\ref{recon2meno}) and (\ref{recon2piu}) belong to accessible portions of the specimen at hand. Our problem is to determine the perturbation $\delta h$ of the  TCC  due to a defect of $\Sigma$. Since the unknown $h$ appears in the left hand side,  reciprocity conditions take the form of a system of equations like
$$F^\pm_{v^\pm}(h,[u]) = G_{a^\pm,\kappa^\pm,\rho^\pm ,c^\pm ,h^\pm,v^\pm,\phi,\psi}$$
that will be written explicitly in the next section\\

\subsection{Suitable families of test functions}\label{sec:test}
We  introduce test functions 
\begin{equation}\label{ftest}
v^\pm(x,z,t)=b_\pm e^{-A t} e^{\mp s_p^\pm z} \cos (px)
\end{equation}
dependent on the parameters $b_\pm$, $A>0$ and  $p=\frac{\pi}{D}k$ ($k=0,1,2,3,...$)  with $s_p^\pm=\sqrt{\frac{A}{\alpha^\pm}+p^2}$ (test functions like these have been introduced in \cite{BAB03} to evaluate an emerging crack from thermal data).
It is easy to check that:\\

(i)  $v^\pm$ vanishes for $t \to \infty$ uniformly in $(x, z) \in \Omega$. More precisely, $\forall \epsilon \in (0,1)$ we have $|v^\pm|<\epsilon $ for $t > \frac{-\ln \epsilon}{A}$;

(ii) $v^\pm_x(x,z,t)= -b_\pm p e^{-A t}e^{\mp s^\pm _p z} \sin(px)$;

(iii) $v^\pm_z(x,z,t)=\mp b_\pm s^\pm_p e^{-A t} e^{\mp s^\pm_p z}  \cos(px)$.\\

The real numbers  $b_+$ and $b_-$ must be determined so that $\kappa^+ v^+_z(x,0,t)=-\kappa^- v^-_z(x,0,t)$ i.e. $\kappa^-b_-s_p^-=\kappa^+b_+s_p^+$ for reasons that will be clear when deriving formula (\ref{rec00}). 
Indeed, we have
\begin{equation}
b_+=1 \phantom{qwerty}; \phantom{qwerty} b_- = \sqrt \frac{A\kappa^+\rho^+c^+ +{\kappa^+}^2 p^2}{A\kappa^-\rho^-c^-+{\kappa^-}^2 p^2}.
\end{equation}

\subsection{Handling  (\ref{recon2meno}) and (\ref{recon2piu}) . Plugging data and test functions}\label{sec:perturb}
{\bf Notation:} (i) $\int_{TD} \equiv \int_0^T \int_{0}^D dt dx$;  (ii) $[u](x,t)= u^-(x,0,t)-u^+(x,0,t)$. \\

\noindent Rewrite (\ref{recon2meno}) as
\begin{equation}\label{recnewmen}
\int_{TD}  v^-(0)h[u]+u^-(0)\kappa^-v^-_z(0) \approx 
\int_{TD}-v^-(-a^-) (h^-u^-(-a^-)-\phi)+u^-(-a^-)\kappa^-v^-_z(-a^-)
\end{equation}
and (\ref{recon2piu}) as  
\begin{equation}\label{recnewpiu}
\int_{TD}  v^+(0)h[u]+u^+(0)\kappa^+v^+_z(0)\approx  \int_{TD}+v^+(a^+) h^+u^+(a^+)+u^+(a^+)\kappa^+v^+_z(a^+) .
\end{equation}
We recall that $\kappa^-v^-_z(0)=-\kappa^+v^+_z(0)$ and we sum (\ref{recnewpiu}) and (\ref{recnewmen}). We get:
\begin{equation}\label{rec00}
\int_{TD} \kappa^-v^-_z(0)[u]+ (v^-(0)+v^+(0)) h[u]  \approx 
N
\end{equation}
where
\begin{equation}\label{rec01}
\begin{split}
N=\int_{TD}-v^+(a^+) h^+u^+(a^+)-u^+(a^+)\kappa^+v^+_z(a^+)-v^-(-a^-) (h^-u^-(-a^-)-\phi)+&\\u^-(-a^-)\kappa^-v^-_z(-a^-)
\end{split}
\end{equation}

Each term in (\ref{rec00}) and (\ref{rec01}) includes $v^\pm$ or $v^\pm_z$ from which  we extract the factor $e^{-At}$ and take the limit for $T \to \infty$ of  integrals $\int_{TD}$. Laplace transform
\begin{equation}
{U}^\pm (x,z) =  \int_0^\infty u^\pm(x,z,t)e^{-A t}dt
\end{equation}
is useful in order to simplify the problem.
In this way, we obtain
\begin{equation}\label{rec00lap}
\int_0^{D}( (b_-+1)h[U](x) + \kappa^+\sqrt{\frac{A}{\alpha^+}+p^2}[U](x)) \cos(px) dx \approx 
\int_0^{D}G^A_p(x)\cos(px)dx
\end{equation}
where $[U]$ was defined in (\ref{diffe}) and
\begin{equation}\label{rec01lap}
G^A_p=U^+(x,a^+)(h^+-\kappa^+s^+_p)e^{-s^+_pa^+}+U^-(x,a^-)b_-(\kappa^-s^--h^-)e^{-s^-a^-}+b_-\Phi e^{-s^-a^-}
\end{equation}
is fully known.\\

\section{Perturbations and inversion formula}\label{sec:pert}

In case of small defects, a perturbation of the interface (and consequelntly of the TTC) is expected. 
More precisely, assume that the TCC of the damaged interface has the form
\begin{equation}
h(x)=h_0+\delta h(x)
\end{equation}
with $\frac{\sup |\delta h|}{h_0}<<1$. We write the solution of  (\ref{equatmL})-(\ref{TC2L}) in the form
\begin{equation}
U^\pm(x,z)=U_0^\pm(x,z)+\delta U^\pm(x,z)
\end{equation}
where the background solution $U_0$ corresponds to the undamaged boundary.  Thanks to stability estimates  of Theorem 3.4 in \cite{IO22}, it is  $\frac{\|\delta U^\pm\|}{\|U^\pm_0\|}$ for suitable norms.
In particular,  following the notation in (\ref{diffe}),
\begin{equation}
[U](x)=[U_0](x)+[\delta U](x).
\end{equation}

We claim the following approximated relations that will be derived in section \ref{linear}:
\begin{equation}\label{claim}
\begin{split}
[\delta U] & \approx -E[U_0] \delta h \\
\delta U^- (x,0) &  \approx   -E^- [U_0] \delta h  \\
\delta U^+(x,0) & \approx   E^+ [U_0]  \delta h.
\end{split}
\end{equation}
with $E^-=\frac{K^-}{1+ (K^-+K^+)h_0}$, $E^+=\frac{K^+}{1+ (K^-+K^+)h_0}$ and $E=E^-+E^+$ where $K^-$ and $K^+$ are explicitely written in (\ref{kappa}). It is easy to get (\ref{claim}) in one dimension  since a solution of (\ref{equatmL})-(\ref{TC2L}) is a linear combination of hyperbolic sine and cosine (see details in  section \ref{linear}). The extension to 2D (or more) is not rigorous but reasonable  as long as the defect (and consequently $\delta h(x)$) is sufficiently smooth. It is remarkable that the constants $E$ and $E^\pm$ depend on the parameter $A$. The choice of the numerical value of $A$ will be a crucial step in the solution of the problem.\\

We apply the perturbative notation, Laplace transform, the approximated relations (\ref{claim}) and, finally, linearization  to the transmission condition (\ref{recnewpiu}) on the positive side of the interface. We recall that here $p=\frac{ \pi}{D}n$ ($n=0,1,2,...$) and $s^+_p= \sqrt{\frac{A}{\alpha^+}+p^2}$. \\

First we write the condition explicitly:
\begin{equation}\label{interpiu}
\begin{split}
\int_{0}^D ( \int_0^T  e^{-At}\cos(px)(h_0 +\delta h(x))([u_0]+[\delta u])dt) dx-&\\
\int_{0}^D ( \int_0^T  (u_0^+(x,0,t)+\delta u^+(x,0,t) \kappa^+ s^+_p e^{-At} \cos(px)dt ) dx \approx  d^+_p
\end{split}
\end{equation}
where
\begin{equation}
\begin{split}
d^+_p =\int_{0}^D ( \int_0^T  e^{-At}e^{-a^+s^+_p  }  \cos(px) h^+(u^+_0(x,a^+,t)+\delta u^+(x,a^+,t)) dt)dx & \\
-\int_{0}^D ( \int_0^T (u^+_0(x,a^+,t)+\delta u^+(x,a^+,t))\kappa^+s^+_p  e^{-At}e^{-a^+s^+_p }  \cos(px) dt)dx.
\end{split} 
\end{equation}
After Laplace transform we have
\begin{equation}\label{interpiuLap}
\int_{0}^D(h_0 +\delta h)([U_0]+[\delta U])  \cos(px) dx-
\kappa^+s^+_p  \int_{0}^D (U_0^+(x,0)+\delta U^+(x,0)) \cos(px) dx \approx  d^+
\end{equation}
where
\begin{equation}
\begin{split}
d^+_p =\int_{0}^D e^{-a^+s^+_p  }  \cos(px) h^+(U^+_0(x,a^+)+\delta U^+(x,a^+))dx & \\
-\int_{0}^D (U^+_0(a^+)+\delta U^+(x,a^+))\kappa^+s^+_p  e^{-a^+s^+_p }  \cos(px) dx.
\end{split} 
\end{equation}
At this point: \\

\noindent (i) subtract the background transmission condition, 

\noindent (ii) normalize $h_0 +\delta h$ into $1+\frac{\delta h}{h_0}$ and $[U_0]+[\delta U]$ into $1+\frac{[\delta U]}{[U_0]}$,

\noindent (iii) neglect second order terms.\\

In this way,  if $\tilde {\delta h} = \frac{\delta h}{h_0}$,  $\tilde {[\delta U]} = \frac{[\delta U]}{[U_0]}$ and  $\tilde {\delta U^+} =\frac{\delta U^+}{[U_0]}$, we get the following linear  equation in the normalized unknowns (observe that $\delta d^+$ is fully known):
\begin{equation}\label{interpiuLpert}
\int_{0}^D(\tilde {\delta h} +\tilde{[\delta U]})  \cos(px) dx-
\frac{\kappa^+s^+_p}{h_0} \int_{0}^D\tilde{ \delta U^+} \cos(px) dx \approx  \tilde{\delta d^+_p}
\end{equation}
where
\begin{equation}
\begin{split}
\tilde{\delta d^+_p} =e^{-a^+ s^+_p } \frac{( h^+
-\kappa^+ s^+_p  )}{h_0} \int_{0}^D \frac{\delta U^+(x,a^+)}{[U_0]} \cos(px) dx.
\end{split} 
\end{equation}
The last step consists in using the claimed relations (\ref{claim}) between   $\delta h$ and $[\delta U]$. Hence, we get ($p=\frac{\pi}{D}n$)
\begin{equation}\label{rec01del}
\int_{0}^D\tilde{\delta h} \cos(px) dx
 \approx \frac{e^{-a^+ s^+_p } ( h^+
-\kappa^+ s^+_p  )}{h_0 (1 -h_0  E -\kappa^+s^+_p  E^+ )   }\int_{0}^D  \delta U^+(x,a^+) \cos(px) dx.
\end{equation}\\

\subsection{The divisor is not zero}\label{sec:stab}

The divisor $1 -h_0  E -\kappa^+s^+_p  E^+   $ requires a discussion. Observe that
$$1 -h_0  E -\kappa^+s^+_p  E^+  =\frac{1-\kappa^+s^+_p K^+}{1+ (K^-+K^+)h_0}.$$

First, consider the term $p=0$ i.e. the mean values of data and unknown. 
Assume that $\epsilon= \frac{\kappa^+s^+_0 }{h^+}$ is greater than $1$  (i.e. $h^+ <\kappa^+ s^+_0$, which is physically very reasonable) and $\sigma =s^+_0 a^+$. We have 
\begin{equation}
\begin{split}
\kappa^+s^+_0 K^+ =  \kappa^+s^+_0 \frac{\kappa^+ s^+_0 \cosh (\sigma)+h^+ \sinh (\sigma)}{\kappa^+ s^+_0 \sinh (\sigma)+h^+ \cosh (\sigma)}\frac{1}{\kappa^+ s^+_0 }\\
= \frac{\kappa^+ s^+_0 \cosh (\sigma)+h^+ \sinh (\sigma)}{\kappa^+ s^+_0 \sinh (\sigma)+h^+ \cosh (\sigma)}\\
=\coth(\sigma) \frac{\epsilon +\frac{1}{\coth(\sigma)}}{\epsilon +\coth(\sigma)} \equiv G_\epsilon(\sigma).
\end{split}
\end{equation}
It is easy to see that, since $\epsilon$ is greater than one, $G_\epsilon(\sigma) > G_1(\sigma)=1$. Since, clearly, $s^+_p>s^+_0>0$ the divisor in (\ref{rec01del}) is different from $0$.



\section{Main result is tested numerically}\label{sec:numer}

The inversion formula (\ref{rec01del}) allows the computation of  an arbitrary number $N$ of cosine-Fourier coefficients  $C_k$ of $\delta h$. Once a value for Laplace parameter $A$ has been chosen,  the solution of the inverse problem defined in section \ref{sec:inverse}  is approximated by the finite sum 
$\sum_{k=0}^{N-1} C_k \cos(\frac{\pi}{D}k x)$ . 
 
\begin{figure}[h]
    \centering
    \begin{subfigure}[b]{0.48\textwidth}
        \centering
        \includegraphics[width=\textwidth]{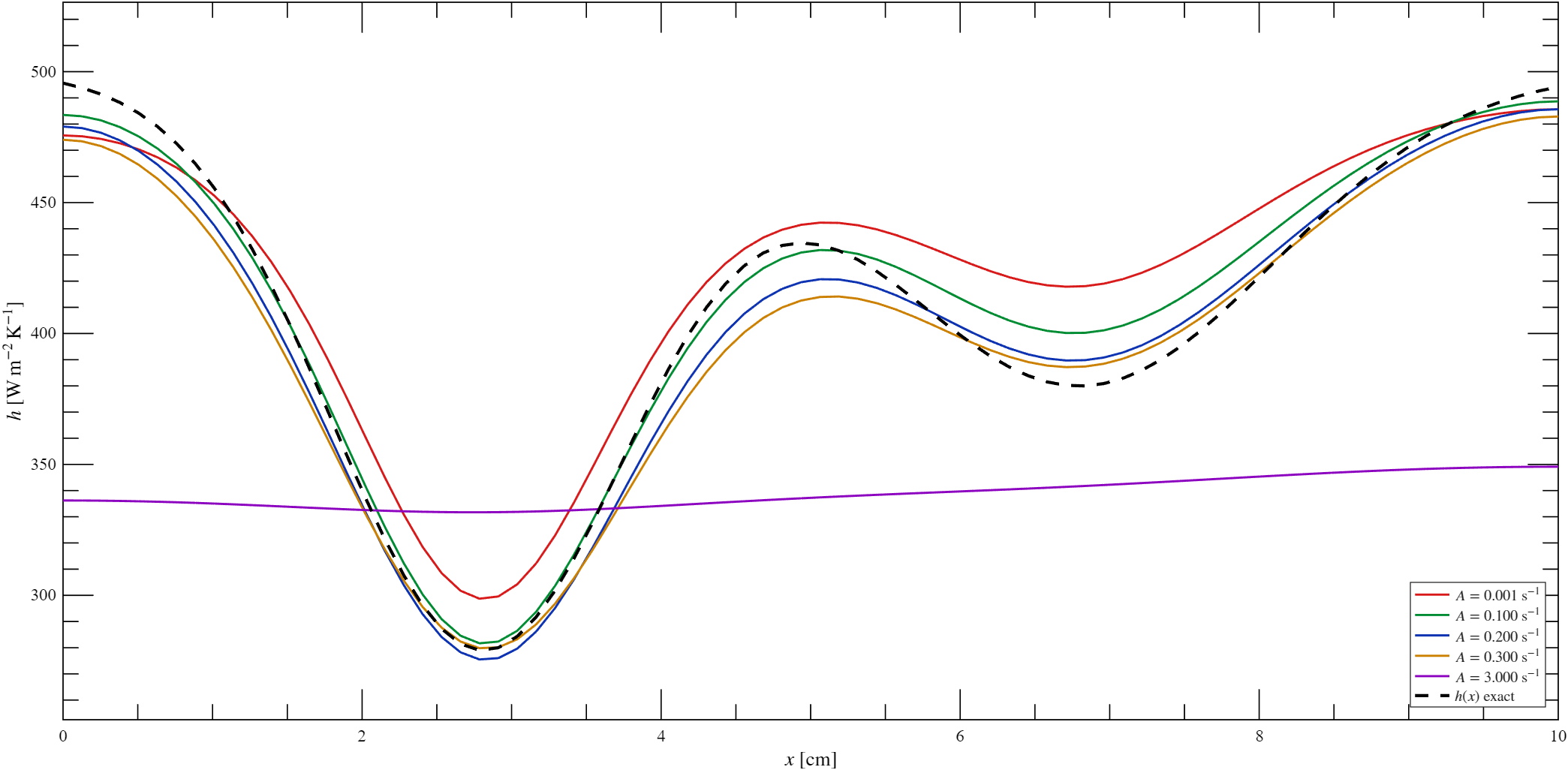}
        \caption{Different values of $A$: Two modes defect}
        \label{fig:small_A}
    \end{subfigure}
    \hfill 
    \begin{subfigure}[b]{0.48\textwidth}
        \centering
        \includegraphics[width=\textwidth]{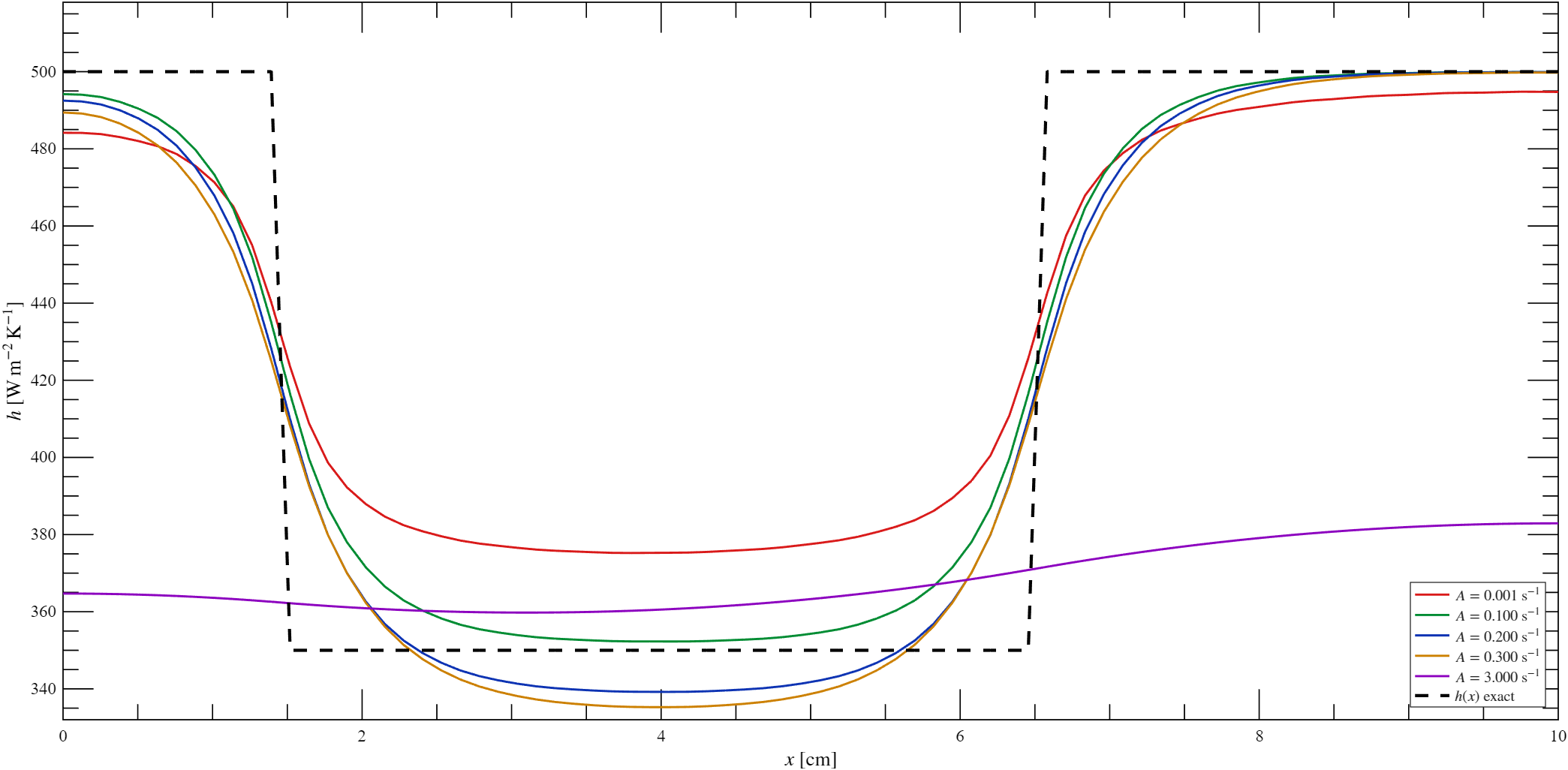}
        \caption{Different values of $A$: Rectangular defect}
        \label{fig:opt_A}
    \end{subfigure}   
    \label{fig:main_label}
\end{figure}

There is a range in which $A$ is meaningful  i.e. Laplace transform helps to define the direct problem. On the other hand, while $A$ is too large (Laplace transform becomes zero) or too small (Laplace transform is simply the time integral of the temperature) Laplace transform is not useful. Here, we adopt the heuristic strategy shown in Figures \ref{fig:small_A} and \ref{fig:opt_A} in order to select $A$. Inversion formula (\ref{rec01del})  is applied to reconstruct the perturbation $\delta h$ of TCC at interface $\Sigma$ corresponding to a range of reasonable  $A$'s (the interval $(.001,3)$ in our tests). A good choice of $A$ is determined  by the fact of identifying objects of similar size and shape so that  an acceptable reconstruction of the unknown $\delta h$ is any of the curves obtained for  $A \in (0.1,0.3)$.

%



\noindent {\bf Remark} Numerical work in this paper is made up of two different steps. First we simulated data collection by solving the direct problem  described in detail in subsection \ref{dirmod} and computing Laplace transform. Then, we implemented formula (\ref{rec01del}) which gives a number of Fourier coefficients of the unknown term $\delta h$.   Numerical analysis is quite standard in both cases. Actually, the goal of this paper consists in deriving formula (\ref{rec01del}) by means of  mathematical analysis instead of applying one of the available numerical methods for the solution of inverse problems or by implementing a new one. 

\section{Linear relation $[\delta U]$ vs $\delta h$ in the one dimensional problem}\label{linear} 

The solutions of (\ref{equatmL})-(\ref{TC2L})  with constant parameters $\Phi$ and $h$ (one dimensional problem in the space variable $z$) take the form
\begin{equation}
U^\pm(z)=B^\pm \cosh (s^\pm_0 z)+C^\pm \sinh (s^\pm_0 z)
\end{equation}
where $s^\pm_0=\sqrt{\frac{A}{\alpha^\pm}}$.
Transmission conditions for $z=0$ determine the values of $B^\pm$ and $C^\pm$.  Since
\begin{equation}
U_z^\pm(z)=B^\pm s^\pm_0 \sinh (s^\pm_0 z)+C^\pm s^\pm_0 \cosh (s^\pm_0 z)
\end{equation}
we have
\begin{equation}\label{cmeno}
C^-=\frac{s^+_0\kappa^+}{s^-_0\kappa^-} C^+
\end{equation}
with
\begin{equation}\label{cpiu}
C^+ =\frac{h(B^+  -B^-)}{\kappa^+ s^+_0 }. 
\end{equation}
We plug (\ref{cmeno}) and (\ref{cpiu}) in the boundary conditions
\begin{equation}
\begin{split}
 \kappa^- B^-  s^-_0 \sinh (s^-_0 a^-)- \kappa^- C^-  s^-_0 \cosh (s^-_0 a^-)+h^-(B^- \cosh (s^-_0 a^-)-C^- \sinh (s^-_0 a^-))&=\Phi\\
 \kappa^+ B^+  s^+_0 \sinh (s^+_0 a^+)+ \kappa^+ C^+  s^+_0 \cosh (s^-_0 a^+)+h^+(B^+ \cosh (s^+_0 a^+)+C^+ \sinh (s^+_0 a^+))&=0\\
\end{split}
\end{equation}
and obtain
\begin{equation}\label{eqin0}
\begin{split}
B^- +h (B^--B^+) K^-&=\frac{\Phi}{\kappa^- s^-_0 \sinh (s^-_0 a^-)+h^- \cosh (s^-_0 a^-)}\\
 B^+-h (B^--B^+)  K^+ &=0
\end{split}
\end{equation}
with
\begin{equation}\label{kappa}
\begin{split}
K^-&=\frac{\kappa^- s^-_0 \cosh (s^-_0 a^-)+h^- \sinh (s^-_0 a^-)}{\kappa^- s^-_0 \sinh (s^-_0 a^-)+h^- \cosh (s^-_0 a^-)}\frac{1}{\kappa^- s^-_0 }\\
K^+ &=\frac{\kappa^+ s^+_0 \cosh (s^+_0 a^+)+h^+ \sinh (s^+_0 a^+)}{\kappa^+ s^+_0 \sinh (s^+_0 a^+)+h^+ \cosh (s^+_0 a^+)}\frac{1}{\kappa^+ s^+_0 }.
\end{split}
\end{equation}
Since
\begin{equation}
[U]=U^-(0)-U^+(0)=B^--B^+,
\end{equation}
we subtract the equations in (\ref{eqin0}) obtaining
\begin{equation}\label{eqin1}
[U]+h[U](K^-+K^+)=\frac{\Phi}{\kappa^- s^-_0 \sinh (s^-_0 a^-)+h^- \cosh (s^-_0 a^-)}.
\end{equation}
In the notation of section \ref{sec:perturb} it is easy to check the background relation
\begin{equation}
[U_0]+h_0[U_0](K^-+K^+)=\frac{\Phi}{\kappa^- s^-_0 \sinh (s^-_0 a^-)+h^- \cosh (s^-_0 a^-)}.
\end{equation}
so that (\ref{eqin1}) becomes
\begin{equation}\label{eqin2}
[\delta U]+\delta h[U_0](K^-+K^+)+h_0[\delta U](K^-+K^+)+O(\delta^2)=0.
\end{equation}
Hence, at the first order, we have
\begin{equation}\label{eqin3}
[\delta U] \approx -\delta h [U_0]\frac{K^-+K^+}{1+ (K^-+K^+)h_0}.
\end{equation}
Since $[\delta U]  = \delta B^--\delta B^+$, it follows from straightforward calculations that
\begin{equation}
\begin{split}
\delta B^-  & =   -\delta h [U_0]\frac{K^-}{1+ (K^-+K^+)h_0}\\
\delta B^+ & =   \delta h [U_0]\frac{K^+}{1+ (K^-+K^+)h_0}.
\end{split}
\end{equation}

{\bf Acknowledgments} Gabriele Inglese would like to thank a long list of scientists that shared with him the study of applied mathematics in this last fourthysix years: Pucci, Talenti, Natalini, Gautschi, Beretta, Bianchi, Ben Abda, Sgheri, Vessella, Francini, Gronchi, Fassari, Pirillo, Franceschini, Gallavotti, Chierchia, Santosa, Mariani, Ceseri, Clarelli, Alves, Olmi, Scalbi, Bison, Alessandrini, Quarteroni.

\end{document}